\documentclass[10pt,twocolumn]{article}
\usepackage{pdfpages}
\usepackage{amssymb}
\usepackage{graphicx}
\usepackage{multirow}
\usepackage{url}
\usepackage{gensymb}
\usepackage{tikz}
\usepackage{amssymb}
\usepackage{amssymb}
\usepackage{pifont}

\usepackage{amsmath}
\usepackage[T1]{fontenc}
\usepackage[utf8]{inputenc}
\usepackage{authblk}
\usepackage{listings}
\usepackage{graphicx}
\usepackage{times}
\usepackage{amssymb}
\usepackage{graphicx}
\usepackage{array}
\usepackage{multirow}
\newcommand{\cov}{\ensuremath\mathrm{Cov}}

\title{Stealing PINs via Mobile Sensors: \\Actual Risk versus User Perception}
\author{Maryam Mehrnezhad, Ehsan Toreini, Siamak F. Shahandashti, Feng Hao}
\affil{School of Computing Science, Newcastle University, Newcastle upon Tyne, UK}

\date{}
\begin{document}
\maketitle

\begin{abstract}
In this paper, we present the actual risks of stealing user PINs by using mobile sensors versus the perceived risks by users. First, we propose PINlogger.js which is a JavaScript-based side channel attack revealing user PINs on an Android mobile phone. In this attack, once the user visits a website controlled by an attacker, the JavaScript code embedded in the web page starts listening to the motion and orientation sensor streams without needing any permission from the user. By analysing these streams, it infers the user's PIN using an artificial neural network. Based on a test set of fifty 4-digit PINs, PINlogger.js is able to correctly identify PINs in the first attempt with a success rate of 74\% which increases to 86\% and 94\% in the second and third attempts respectively. 
The high success rates of stealing user PINs on mobile devices via JavaScript indicate a serious threat to user security. 

With the technical understanding of the information leakage caused by mobile phone sensors, we then study users' perception of the risks associated with these sensors. We design user studies to measure the general familiarity with different sensors and their functionality, and to investigate how concerned users are about their PIN being discovered by an app that has access to all these sensors. Our studies show that there is significant disparity between the actual and perceived levels of threat with regard to the compromise of the user PIN. We confirm our results by interviewing our participants using two different approaches, within-subject and between-subject, and compare the results. We discuss how this observation, along with other factors, renders many academic and industry solutions ineffective in preventing such side channel attacks. 

\end{abstract}
\section{Introduction}
\label{Intro}
Smartphones equipped with many different sensors such as \textit{GPS, light, orientation and motion} are continuously providing more features to end users in order to interact with their real-world surroundings. Developers can have access to the mobile sensors either by 
1) writing native code using mobile OS APIs~\cite{AndDev}, 
2) recompiling HTML5 code into a native app~\cite{CodeInjection}, or 
3) using standard APIs provided by the W3C which are accessible through JavaScript code within a mobile browser\footnote{w3.org/TR/\#tr\_Javascript\_APIs}. 
The last method has the advantage of not needing any app-store approval for releasing the app or doing future updates. More importantly, the JavaScript code is platform independent, i.e., once the  code is developed it can be executed within any modern browser on any mobile OS. 

\textbf{In-browser access risks.} While sensor-enabled mobile web applications provide users more functionalities, they raise new privacy and security concerns. 
Both the academic community and the industry have recognised such issues regarding certain sensors such as geolocation \cite{GPS}. For a website to access the geolocation data, it must ask for explicit user permission. However, to the best of our knowledge, there is little work evaluating the risks of in-browser access to other sensors. Unlike in-app attacks, an in-browser attack, i.e., via JavaScript code embedded in a web page, does not require any app installation. In addition, JavaScript code does not require any user permission to access sensor data such as device motion and orientation. Furthermore, there is no notification while JavaScript is reading the sensor data stream. Hence, such in-browser attacks can be carried out far more covertly than the in-app counterparts. 

However, an effective in-browser attack still has to overcome the technical challenge that the sampling rates available in browser are much lower than those in app. For example, as we observed in \cite{Mehrnezhad}, frequency rates of motion and orientation sensor data available in-browser are 3 to 5 times lower than those of accelerometer and gyroscope available in-app. 
\begin{figure*}[t]
	\centering
	\includegraphics[scale = 0.15]{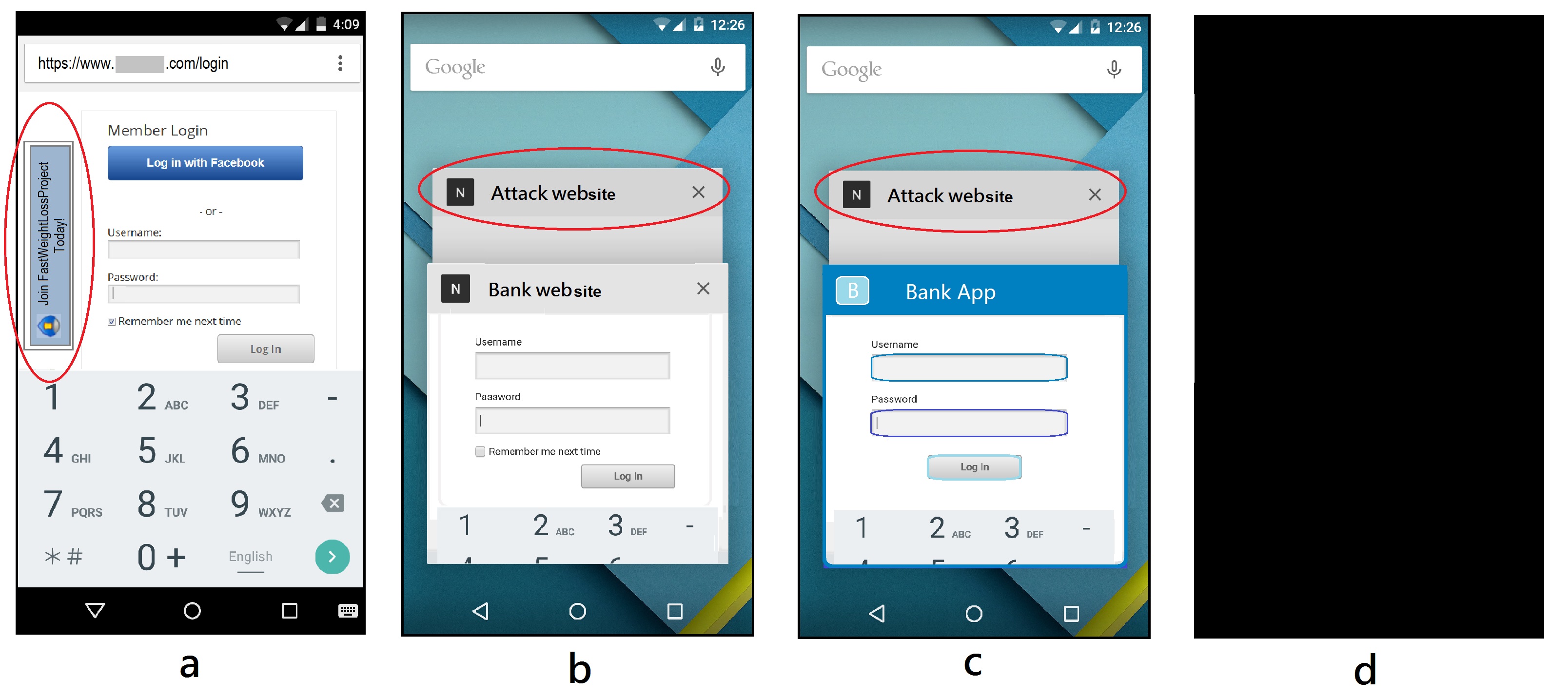}
	\caption{ 
	PINlogger.js potential attack scenarios; a) the malicious code is loaded in an iframe and the user is on the same tab, b) the attack tab is already open and the user is on a different tab, c) the attack content is already open in a minimised browser, and the user is on an installed app, d) the attack content is already open in a (minimised) browser, and the screen is locked.  
	The attacker listens to the side channel motion and orientation measurements of the victim's mobile device through JavaScript code, and uses machine learning methods to discover the user's sensitive information such as activity types and PINs.
	}  
		\label{attack}
\end{figure*}

\textbf{In-browser attacks.} Many popular browsers such as Safari, Chrome, Firefox, Opera and Dolphin have already implemented access to the above sensor data. As we demonstrated in \cite{TouchSig} and \cite{Mehrnezhad}, all of these mobile browsers allow such access when the code is placed in any part of the active tab including \textit{iframes} (Figure \ref{attack}, a). In some cases such as Chrome and Dolphin on iOS, an inactive tab can have access to the sensor measurements as well (Figure \ref{attack}, b). Even worse, some browsers such as Safari allow the inactive tabs to access the sensor data, when the browser is minimised (Figure \ref{attack}, c), or even when the screen is locked (Figure \ref{attack}, d). 

Through experiments, we find that mobile operating systems and browsers do not implement consistent access control policies in regard to mobile orientation and motion sensor data. Partly, this is because W3C specifications~\cite{W3CMotion} do not specify any policy and do not discuss any risks associated with this potential vulnerability. Also, because of the low sampling rates available in browser, the community have been neglecting the security risks associated with in-browser access to such sensor data. However, in TouchSignatures~\cite{Mehrnezhad}, we showed that despite the low sampling rates, it is possible to identify user touch actions such as click, scroll, and zoom and even the numpad's digits. In this paper, we introduce PINLogger.js, an attack on full 4-digit PINs as opposed to only single digits in~\cite{Mehrnezhad}.

\textbf{Mobile sensors.}
Today, sensors are everywhere: from your personalised devices such as mobiles, tablets, watches, fitness trackers, and other wearables, to your TV, car, kitchen, home, and to the roads, parking lots, and smart cities. These new technologies are equipped with many different sensors such as NFC, accelerometer, orientation and motion and are connected to each other. 
These sensors are continuously providing
more features to end users in order to interact with their real world
surroundings. While the users are benefiting from richer and more personalised apps which are using these sensors for different applications such as fitness, gaming, and even security application such as authentication, 
the growing number of sensors introduces new security and privacy risks to end users, and makes the task of sensor management more complex.

\textbf{Research questions.} While sensors on mobile platforms are getting more powerful, and starting to collect more information about the users and their environment, we want to evaluate the general knowledge about these sensors among the mobile users. We are particularly interested to know the level of concern people may have about these sensors being able to threaten their privacy and security.  

\textbf{Contributions.} In this work, we contribute to the study of sensors and their actual risks and their perceived risks by users as follows:

\begin{itemize}
\item We introduce PINLogger.js, an attack on full 4-digit PINs as opposed to only single digits in~\cite{Mehrnezhad}. We show that unregulated access to these sensors impose more serious security risks to the users in comparison with more well-known sensors such as camera, light and microphone. 

\item We conduct user studies to investigate users' understanding about these sensors and also their perception of the security risks associated with them. We show that users in fact have fewer security concerns about these sensors comparing to more well-known ones. 

\item We study and challenge current suggested solutions, and discuss why our studies show they cannot be effective. We argue that a usable and secure solution is not straightforward and requires further research. 
\end{itemize}
\section{User activities} 
\label{activity}
\begin{figure*}[t]
	\centering
	\includegraphics[scale = 0.25]{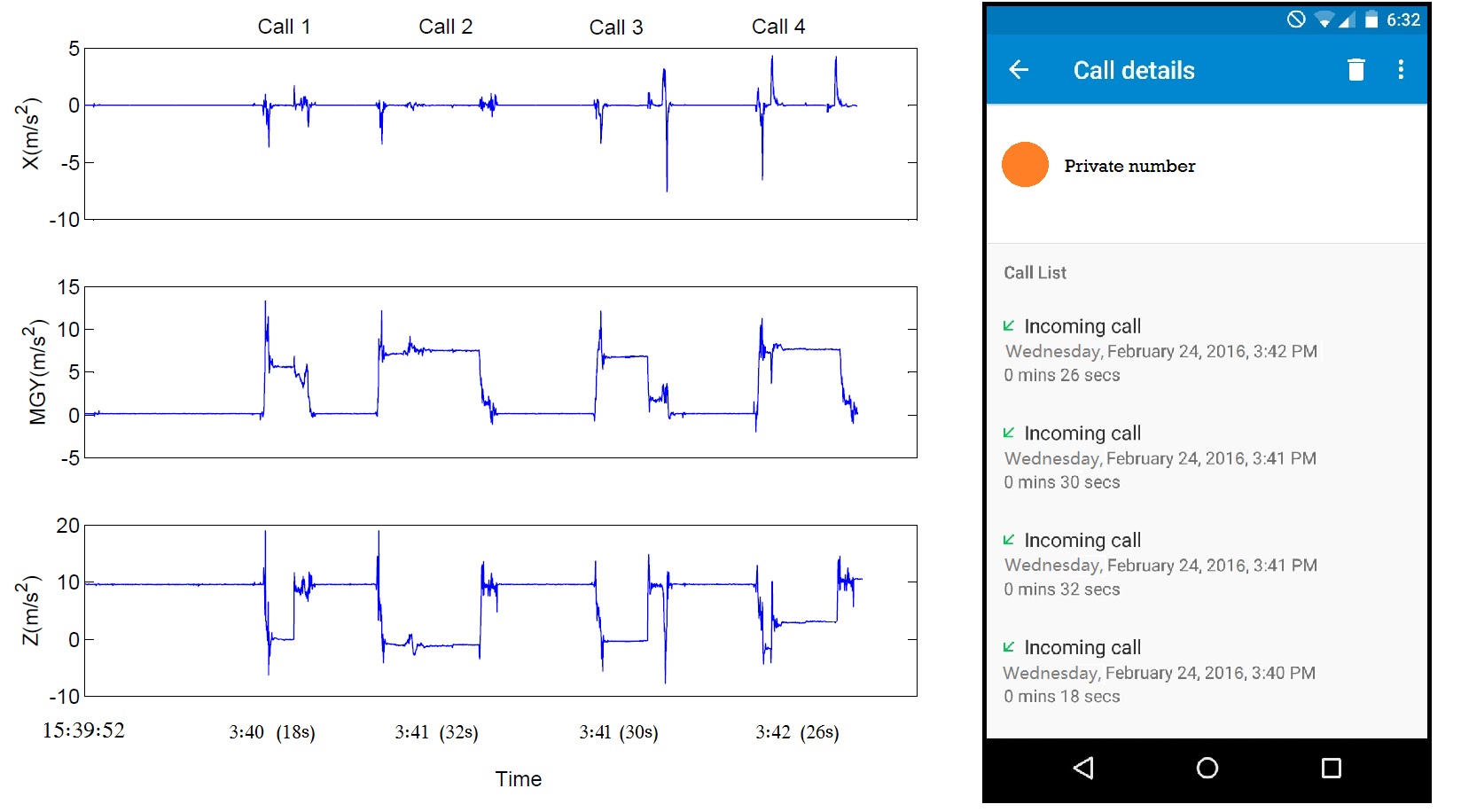}
	\caption{Left: Three dimensions (x, y, and z) of acceleration data including gravity (from the motion sensor). The start time, duration, and end time of four phone calls are easily recognisable from these measurements.
		Right: The screenshot of the call history of the phone during the experiment.}
	\label{call}
\end{figure*}
The potential threats to the user security posed by an unauthorised access to the motion and orientation sensor data are not immediately clear. 
Here we demonstrate two simple scenarios which show that
sensitive user information such as phone calls timing and physical activities can be deduced from device orientation and motion sensor data obtained from JavaScript.

Users tend to move their mobile devices in distinctive manners while performing certain tasks on the devices, or by simply carrying them. 
Examples of the former include answering a call or taking a photo, while the latter covers their transport mode. 
In both cases, an identifiable succession of movements is exhibited by the device. 
As a result, a web-based program which has 
access to the device orientation and motion data 
may reveal sensitive facts about users such as the exact timing information of the start and end of phone calls and that of taking photos. 
On the other hand, while the user is simply carrying her device, the device movement pattern may reveal information about the user's transport mode, e.g., if the user is stationary at one place, walking, running, on the bus, in a car, or on the train. 
We present the results of two initial experiments that we have performed on a Nexus~5 using \textit{Maxthon Browser} (as an example of a browser that allows JavaScript to access sensor data even when the screen is locked).

\textbf{Motion and orientation sensors detail.} Before, presenting the results, we first explain the motion and orientation sensors in detail. According to W3C specifications~\cite{W3CMotion} motion and orientation sensor data are a series of different measurements as follows:
\begin{itemize}
\item device \emph{orientation} which provides the physical orientation of the device, expressed as three rotation angles ($\alpha$, $\beta$, $\gamma$) in the device's local coordinate frame,

\item device \emph{acceleration} which provides the physical acceleration of the device, expressed in Cartesian coordinates ($x$, $y$, $z$) in the device's local coordinate frame,

\item device \emph{acceleration-including-gravity} which is similar to acceleration except that it includes gravity as well,

\item device \emph{rotation rate} which provides the rotation rate of the device about the local coordinate frame, expressed as three rotation angles ($\alpha$, $\beta$, $\gamma$), and

\item \emph{interval} which provides the constant sampling rate 
and is expressed in milliseconds (ms).
\end{itemize}

The device coordinate frame is defined with respect to the standard position of the mobile screen. When it is in the portrait mode, $x$ and $y$ axes are in the plane of the screen and are positive towards the screen's right and up, and $z$ is perpendicular to the plane of the screen and is positive outwards from the screen.  
Moreover, the sensor data discussed above are processed sensor data obtained from multiple physical sensors such as gyroscope and accelerometer. 
In the rest of this paper, unless specified otherwise, by sensor data we mean the sensor data accessible through mobile browsers which include acceleration, acceleration-including-gravity, rotation rate, and orientation.  

\textbf{Phone call timing.}
In the first experiment, we opened the website carrying our Javascript code, and then locked the screen. The Javascript code continued to log orientation and motion data while the Android phone was left on a desk. For this experiment, we used another phone to call the Android phone four times with a few seconds apart between the calls. As demonstrated in Fig.~\ref{call}~(left), the 4 distinct phone calls along with their timing are recognisable from the three dimensions of acceleration (including gravity) which come from the device motion sensor. 
For a better comparison, Fig.~\ref{call}~(right) shows the received call history of the phone during the experiment with their start times and durations. 
As shown in this figure, the captured sensor data match the call history. 

\begin{figure}[t]
	\centering
	\includegraphics[scale = 0.25]{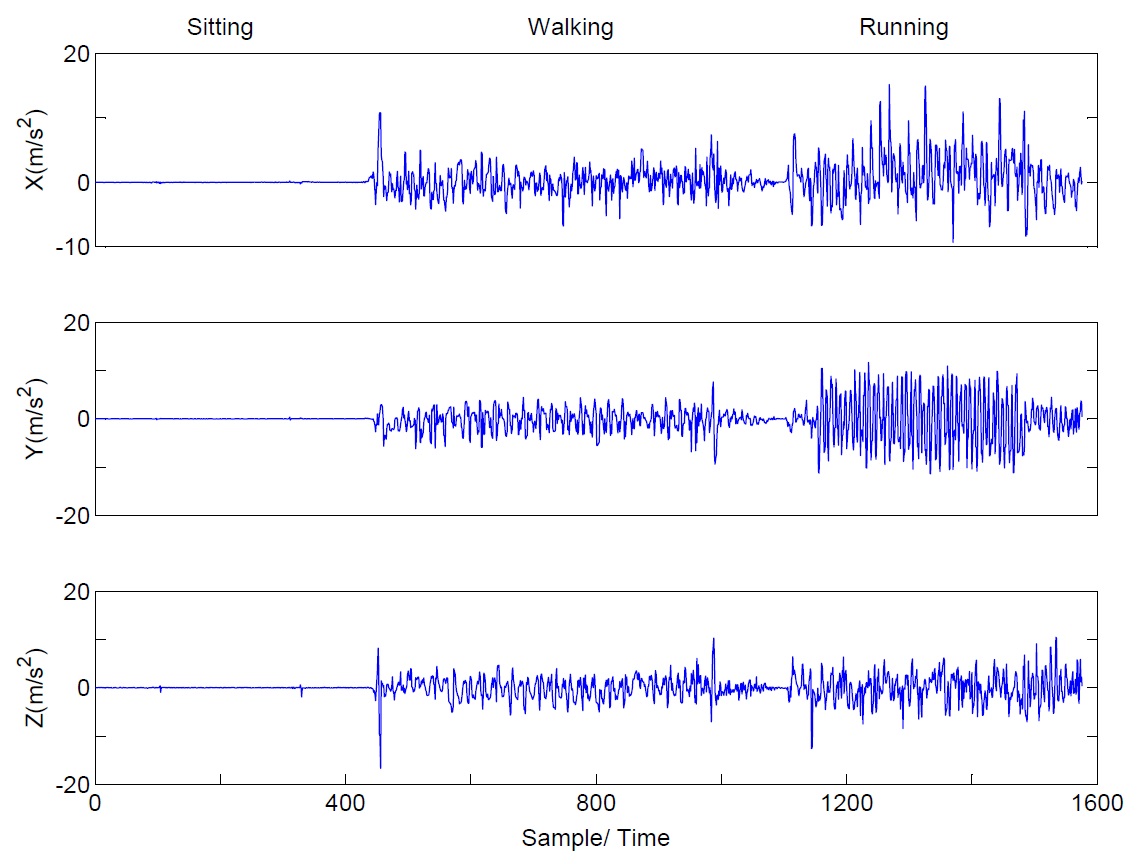}
	\caption{Three dimensions (x, y, and z) of acceleration data (from the motion sensor) during 22 s of sitting, 34 s of walking and 25 s of running.} 
	\label{sitting}
\end{figure}
\textbf{User physical activities.}
In the second experiment, we again locked the phone and recorded the sensor data during 22 seconds of sitting, 34 seconds of walking and 25 seconds of slow running. We observed that the mentioned activities have visibly distinctive sensor streams. As an example, Fig.~\ref{sitting} shows the acceleration data from motion sensor. As it can be seen, the mentioned activities are recognisable from each other since they are visibly different in the sensor measurements.  

Our initial evaluations suggest that discovering device movement related information such as call times and user's mode of transport can be easily implemented. 
However, as we will explain, distinguishing user PINs is a lot harder as the induced sensor measurements are only subtly different. In the following sections, we will demonstrate that, with advanced machine learning techniques, we are able to remotely infer the entered PINs on a mobile phone with high accuracy.

\section{PINlogger.js}
\label{logger}
In this section, we describe an  advanced attack on user's PINs by introducing PINlogger.js. In the following subsections, we describe the attack approach, our program implementation, data collection, feature extraction, and neural network. 

\subsection{Attack approach}
We consider an attacker who wants to learn the user's PIN tapped on a soft keyboard of a smartphone via side channel information. We consider (digit-only) PINs since they are popular credentials used by users for many purposes such as unlocking phone, SIM PIN, NFC payments, bank cards, other banking services, gaming, and other personalised applications such as healthcare, insurance, etc.  
Unlike similar works which have to gain the access through an installed app \cite{Speech:Gyr,accessory,Tapprints,touchlogger,PINCamera,SkimLight,KeyMic,taplogger,Tapprints2,Motionattack}, our attack does not require any user permission. Instead, we assume that the user has loaded the malicious web content in the form of an iframe, or another tab while working with the mobile browser as shown in Figure \ref{attack}. At this point, the attack code has already started listening to the sensor sequences from the user's interaction with the phone. 

In order to uncover when the user enters his PIN, we need to classify his touch actions such as click, scroll, and zoom. We  have already shown in TouchSignatures \cite{Mehrnezhad} that with the same sensor data and by applying classification algorithms, it is possible to effectively identify user's touch actions. Here, we consider a scenario after the touch action classification. In other words, our attacker already knows that the user is entering his PIN.
Moreover, unless explicitly noted, we consider a generic attack scenario which is not user-dependant. This means that we do not need to train our machine learning algorithm with the same user as the subject of the attack. Instead, we have a one-round training phase with data from multiple voluntary users, and use the obtained trained algorithm to output other users' PINs later. This approach has the benefit of not needing to trick individual users to collect their data for training. 

\subsection{Web program implementation}
\begin{figure}[t]
	\centering
	\includegraphics[scale = 0.075]{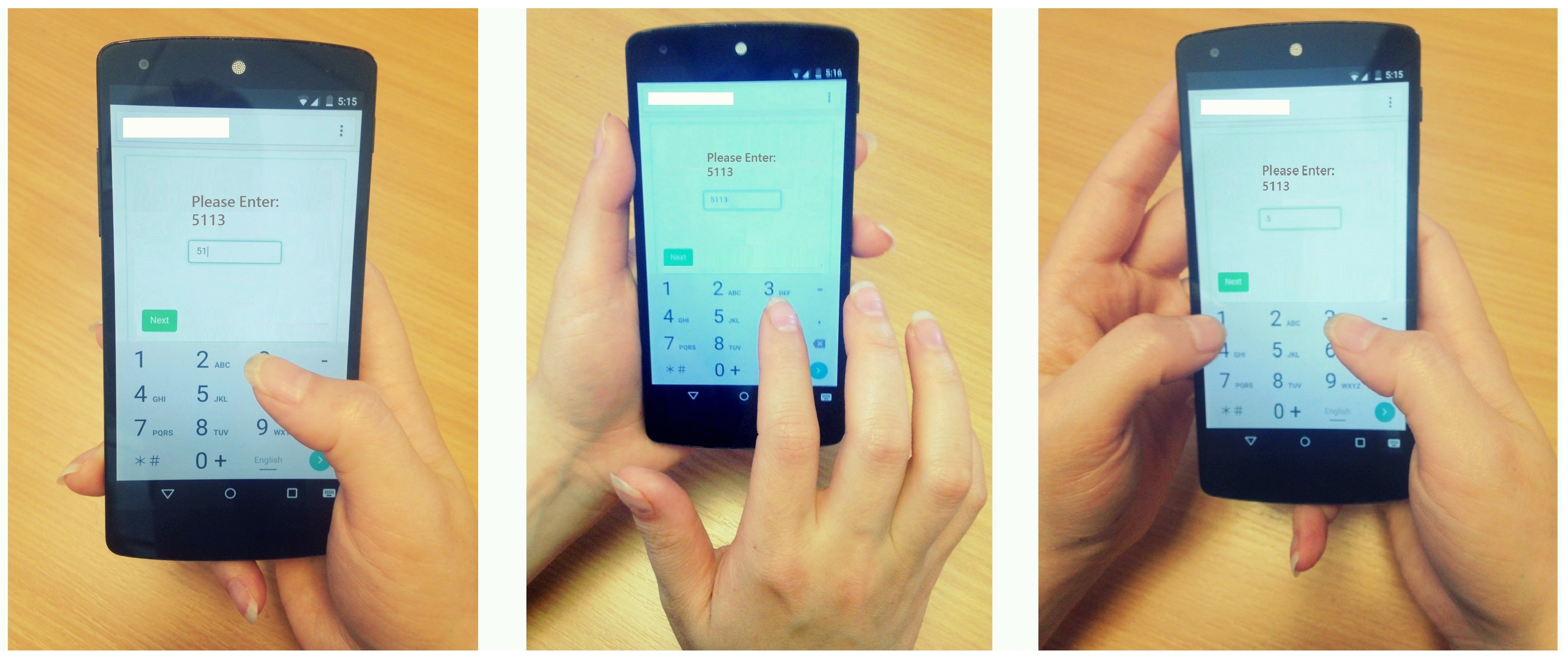}
	\caption{Different input methods used by the users for PIN entrance.}
	\label{Inputmethods}
\end{figure}

We implemented a web page with embedded JavaScript code in order to collect the data from voluntary users. Our code registers two listeners on the window object to have access to orientation and motion data separately.  
The event handlers defined for these purposes are named \textit{DeviceOrientationEvent} and \textit{DeviceMotionEvent} respectively. 
On the client side, we developed a GUI in HTML5 which shows random 4-digit PINs to the users and activates a nummpad for them to enter the PINs as shown in Figure \ref{Inputmethods}. 
All sensor sequences are sent to the database along with their associated labels which are the digits of the entered PINs. 
We implemented our server program using Node.js (nodejs.org). Our code sends the orientation and motion sensor data of
the mobile device to our NoSQL database using MongoLab (mongolab.com, web-based service for MongoDB). 
When the event listener fires, it establishes a socket by using Socket.IO (socket.io) between the client and the server and constantly transmits the sensor data to the database. Both Node.js and MongoDB (as a document-oriented database) 
are known for being capable of supporting data intensive applications in real time. 

In the proof-of-concept implementation of the attack, we focus on working with active web pages, which allows us to easily identify the start of a touch action through the JavaScript access to the \textit{onkeydown} event. 
A similar approach is adopted in other works: e.g., TouchLogger
\cite{touchlogger} and TapLogger \cite{taplogger}. In an extended attack scenario, a more complex segmentation process would be needed to identify the start and end of a touch action. This could be achieved by measuring the peak amplitudes of a signal, as done in \cite{KeyMic}. 

\subsection{Data collection}
Following the approach of Aviv et al.~\cite{Tapprints2} and Spreitzer \cite{SkimLight}, we consider a set of 50 fixed PINs with uniformly distributed digits. We created these PINs in a way that all digits are repeated about the same time (around 20 times). The data collection code is publicly available via github. Technical details of the data collection process and the collected data are publicly available too\footnote{github.com/maryammjd/Reading-sensor-data-for-fifty-4digit-PINs}. 

We conducted our user studies using Chrome on an Android device (Nexus 5). The experiments and results are based on the collected data from 10 users, each entering all the 50 4-digit PINs for 5 times. 
Our voluntary participants were university students and staff and performed the experiments at university offices. We simply explained to them that all they needed was to enter a few PINs shown in a web page. 

In relation to the environmental setting for the data collection, 
we asked the users to remain sitting in a chair while working with the phone. We did not require our users to hold the phone in any particular mode (portrait or landscape) or work with it by using any specific input method (using one or two hands). 
We let them choose their most comfortable posture for holding the phone and working with it as they do in their usual manner. While watching the users during the experiments, we noticed that all of our users used the phone in the portrait mode by default. Users were either leaning their hands on the desk or freely keeping them in the air. We also observed the following input methods used by the users. 
\begin{itemize}
\item Holding the phone in one hand and entering the PIN with the thumb of the same hand (Figure \ref{Inputmethods}, left).
\item Holding the phone in one hand and entering the PIN with the fingers of the other hand (Figure \ref{Inputmethods}, centre). 
\item Holding the phone with two hands and entering the PIN with the thumbs or fingers of both hands (Figure \ref{Inputmethods}, right).
\end{itemize}  

In the first two cases, users exchangeably used either their right hands or left hands in order to hold the phone. In order to simulate a real world data collection environment, we took the phone to each user's workspace and briefly explained the experiment to them, and let them complete the experiment without our supervision. All users found this way of data collection very easy and could finish the experiments without any difficulties. Our participants were given each an Amazon voucher (worth \pounds 10) at the end for their participation.

\subsection{Feature extraction}
In order to build the feature vector as the input to our classifier algorithm, we consider both time domain and frequency domain features. We improve our suggested feature vectors in \cite{Mehrnezhad} by adding some more complex features such as the correlation between the measurements. This addition 
improves the results, as we will discuss in Section \ref{Eva}.
As discussed before, 12 different sequences obtained from the collected data include
orientation (ori), acceleration (acc), acceleration-including-gravity (accG), and rotation rate (rotR)
with three sequences (either x, y and z, or $\alpha$, $\beta$ and $\gamma$) for each sensor measurement. 
As a pre-processing step and in order to remove the effect of the initial position and orientation of the device, we subtract the initial value in each sequence from subsequent values in the sequence. 

We use these pre-processed sequences for feature extraction in time domain directly. In frequency domain, we apply the Fast Fourier transform (FFT) on the pre-processed sequences and use the transformed sequences for feature extraction.  
In order to build our feature vector, first we obtain the maximum, minimum, and average values of each pre-processed and FFT sequences. These statistical measurements give us $3\times 12 = 36$ features in the time domain, and the same number of features in the frequency domain.
We also consider the total energy of each sequence in both time and frequency domains calculated as the sum of the squared sequence values, i.e., $E=\sum{v_i^2}$ which gives us 24 new features. 

The next set of features are in time domain and are based on the correlation between each pair of sequences in different axes. We have 4 different sequences; ori, acc, accG, and rotR, each represented by 3 measurements. Hence, we can calculate 6 different correlation values between the possible pairs; (ori, acc), (ori, accG), (ori, rotR), (acc, accG), (acc, rotR), and (accG, rotR), each presented in a vector with 3 elements. We use the Correlation coefficient function in order to calculate the similarity rate between the mentioned sequences. The correlation coefficient method is commonly used to compare the similarity of the shapes of two signals (e.g. \cite{Shake4}). 
Given two sequences $A$ and $B$ and $\cov(A,B)$ denoting covariance between $A$ and $B$, the correlation coefficient is computed as below: 
\begin{equation}
{R_{AB}} = {\cov(A,B) \over \sqrt{\cov(A,A) \cdot \cov(B,B)}}
\label{eqTime}
\end{equation}

The correlation coefficient of two vectors measures their linear dependence by using  covariance.  
By adding these new 18 features, our feature vector consists of a total of 114 features. 

\subsection{Neural network}
We apply a supervised machine learning algorithm by using an Artificial Neural Network (ANN) to solve this classification problem. 
The input of an ANN system could be either raw data, or pre-processed data from the samples. In our case, we have preprocessed our samples by building a feature vector as described before. Therefore, as input, our ANN receives a set of 114 features for each sample.   
As explained before, we collected 5 samples per each 4-digit PIN from 10 users. While reading the records, we realised that some of the PINs have been entered wrongly by some users. This was expected since each user was required to enter 250 PINs. 
Since we recorded both expected and entered PINs in our data collection, we could easily identify these PINs and exclude them from our analysis. Overall, out of 2500 records collected from 10 users, 12 of the PINs were entered wrongly. Hence we ended up with 2488 samples for our ANN. 

The feature vectors are mapped to specific labels from a finite set: i.e., 50 fixed random 4-digit PINs. 
We train and validate our algorithm with two different subsets of our collected data, and test the neural network against a separate subset of the data. 
We train the network with 70\% of our data, validate it with 15\% of the records and test it with the remaining 15\% of our data set. We use a pattern recognition/classifying network in Matlab with one hidden layer and 1000 nodes. 
Pattern recognition/classifying networks normally use a scaled conjugate gradient (SCG) back-propagation algorithm for updating weight and bias values in training. Scaled conjugate gradient is a fast supervised learning algorithm~\cite{ANNbook}. 

\section{Evaluation}
\label{Eva}
In this section we present the results of our attack on 4-digit PINs in two different forms: multi-users mode, and same-user mode. We also train separate ANN systems to learn individual digits of PINs and compare these results with other works.   

\subsection{Multiple-users mode}
\label{rate}
The second column of Table~\ref{allusers} shows the accuracy of our ANN  trained with the data from all users. In this mode, 
the results are based on training, validating, and testing our ANN using the collected data from all of our 10 participants. 
As the table shows, in the first attempt PINlogger.js is able to infer the user's 4-digit PIN correctly with accuracy of 74.43\%, and as expected it gets better in further attempts.  
By comparison, a random attack can guess a PIN from a set of 50 PINs with the probability of 2\% in the first attempt, and 6\% in three attempts.

\begin{table}[t]
\centering
\begin{tabular}{|l|c|c|}
\hline
Attempts  
& Multiple-users & Same-user \\
\hline
One & 74\% &79\%  \\
Two &  86\% &93\% \\
Three & 94\% &97\% \\
\hline
\end{tabular}
\caption {PINlogger.js's PIN identification rates in different attempts.}
\label{allusers}
\end{table}

\subsection{Same-user mode}
\label{ind}
In order to study the impact of individual training, 
we trained, validated and tested the network with the data collected from one user. 
We refer to this mode of analysis as the same-user mode. 
We asked our user to enter 50 random PINs, each five times, and repeated the experiment for 10 times (rounds). 
The reason we have repeated the experiments is that the classifier needs to receive enough samples to be able to train the system. 
Interestingly, our user used all three different input methods shown in Figure~\ref{Inputmethods} during the PIN entrance. 
As expected, our classifier performs better when it is personalized: the accuracy reaches 79.23\% in the first attempt, and increases to 93.52\% and 97.71\% in two and three attempts, respectively.  

In the same-user mode, convincing the users to provide the attacker with sufficient data for training customised classifiers is not easy, but still possible. Approaches similar to gaming apps such as Math Trainer\footnote{play.google.com/store/apps/details?id=com.solirify.mathgame} could be applied. Math-based CAPTCHAs are possible web-based alternatives.  
Any other web-based game application which segments the GUI similar to a numerical keypad will do as well. Nonetheless, in this paper we mainly follow the multiple-users approach. 

\subsection{Identification of PIN digits}
One might argue that the attack should be evaluated against the whole 4-digit PIN space. However, we believe that the attack could still be practical when selecting from a limited set of PINs since users do not select their PINs randomly~\cite{randomPINs}. It has been reported that around 27\% of all possible 4-digit PINs belong to a set of 20 PINs\footnote{datagenetics.com/blog/september32012/}, including straightforward ones like `1111', `1234', or `2000'. 
Nevertheless, we present the results of our analysis of the attack against the entire search space for the two experiment modes discussed above. We considered 10 classes of the entered digits (0--9) from the data we collected on 4-digit PINs used in Section~\ref{rate}.

In the multiple-users mode, we trained, validated, and tested our system with data from all 10 users. In the same-user mode, we trained personalised classifiers for each user. 
Unlike the test condition of Section~\ref{ind}, we did not have to increase the number of rounds of PIN entry here since we had enough samples for each digit per user. In the same-user mode in this section, we used the average of the results of our 10 users.  
The average identification rates of different digits for three different approaches are presented in Table \ref{allusersDigits}.
\begin{table}[t]
\centering
\begin{tabular}{|l|c|c|}
\hline
Attempts  
& Multiple-users & Same-user \\ 
\hline
One & 70\% & 79\% \\
Two & 83\% & 90\%\\
Three & 92\% & 96\% \\
\hline
\end{tabular}
\caption {Average digit identification rates in different attempts.}
\label{allusersDigits}
\end{table}

The results in our multiple-users mode indicate that we can infer the digits with a success probability of 70.75\%, 83.27\% and 92.06\% in the first, second, and third attempts, respectively. 
This means that for a 4-digit PIN and based on the obtained sensor data, the attacker can guess the PIN from a set of $3^4 = 81$ possible PINs with a probability of success of $0.9206^4 = 71.82\%$. 
A random attack, however, can only predict the 4-digit PIN with the probability of 0.81\% in 81 attempts. By comparison, PINlogger.js achieves a dramatically higher success rate than a random attacker.

\begin{table*}[!t]
\centering
\begin{tabular}{|l|lll|lll|}
\hline
Features & Sensor & Access & Training & \multicolumn{3}{c|}{Identification rate} \\
Work&&  &  & 1st try & 2nd try & 5th try\\
\hline
PIN Skimming \cite{SkimLight} & Light& in-app & same-user& NA& 50\%& 65\%\\

PIN Skimmer \cite{PINCamera}& Cam, Mic& in-app& same-user & NA& 30\%& 50\%\\

Keylogging by Mic \cite{KeyMic} & Mic, Gyr& in-app & same-user& 94\%& NA & NA \\

TapLogger \cite{taplogger} & Acc, Ori& in-app& same-user& 40\%& 75\%& 100\%\\

Acc side channel \cite{Tapprints2} & Acc& in-app& same-user& 18\%& NA& 43\%\\

\hline
PINlogger.js &Motion, Ori& in-browser& multiple-users& 74\%& 86\%& 98\%\\
&& & same-user& 79\%& 93\%& 99\%\\
\hline
\end{tabular}
\caption {Comparison of PINlogger.js with related attacks on 4-digit PINs.} 
\label{compareall}
\end{table*}
Using a similar argument, in the same-user mode the success probability of guessing the PIN in 81 attempts is 85.46\%.
In the same setting, Cai and Chen report a success rate of 65\% using accelerometer and gyroscope data~\cite{Keystrokes} and Simon and Anderson's PIN Skimmer only achieves a 12\% success rate in 81 attempts using camera and microphone~\cite{PINCamera}.     
Our results in digit recognition in this paper are also better than what is achieved in TouchSignatures~\cite{Mehrnezhad}. 
In summary, PINlogger.js performs better than all sensor-based digit-identifier attacks in the literature. 

\subsection{Comparison with related work}
\label{comp}
Obtaining sensitive information about users such as PINs based on mobile sensors has been actively explored by researchers in the field \cite{WiFiattack, smartwatch}. 
In particular, there is a number of research which use mobile sensors through a malicious app running in the background to extract PINs entered on the soft keyboard of the mobile device. 
For example, GyroPhone, by Michalevsky et al.~\cite{Speech:Gyr}, shows that gyroscope data is sufficient to identify the speaker and even parse speech to some extent. 
Other examples include Accessory \cite{accessory} by Owusu et al.~and Tapprints \cite{Tapprints} by Miluzzo. They infer passwords on full alphabetical soft keyboards based on accelerometer measurements. Touchlogger \cite{touchlogger} is another example by Cai and Chen \cite{Keystrokes} which shows the possibility of distinguishing user's input on a mobile numpad by using accelerometer and gyroscope. The same authors demonstrate a similar attack in \cite{Motionattack} on both numerical and full keyboards. The only work which relies on in-browser access to sensors to attack a numpad is our previous work, TouchSignatures \cite{Mehrnezhad}. All of these works, however, aim for the individual digits or characters of a keyboard, rather than the entire PIN or password.   

Another category of works directly target user PINs. For example, PIN skimmer by Simon and Anderson~\cite{PINCamera} is an attack on a user's numpad and PINs using the camera and microphone on the smartphone. Spreitzer suggests another PIN Skimming attack~\cite{SkimLight} and steals a user's PIN based on the measurements from the smartphone's ambient light sensor. Narain et al.\ introduce another attack~\cite{KeyMic} on smartphone numerical and alphabetical keyboards and the user's PINs and credit card numbers by using the smartphone microphone. TapLogger by Xu et al.~\cite{taplogger} is another attack on the smartphone numpad which outputs the pressed digits and PINs based on accelerometer and orientation sensor data. Similarly, Aviv et al.\ introduce an accelerometer-based side channel attack on the user's PINs and patterns in~\cite{Tapprints2}. We choose to compare PINlogger.js with the works in this category since they have the same goal of revealing the user's PINs. Table \ref{compareall} presents the results of our comparison. 

As shown in Table \ref{compareall}, PINlogger.js is the only attack on PINs which acquires the sensor data via JavaScript code. In-browser JavaScript-based attacks impose even more security threats to users since unlike in-app attacks, they do not require any app installation and user permission to work. Moreover, the attacker does not need to develop different apps for different platforms such as Android, iOS, and Windows. Once the attacker develops the JavaScript code, it can be deployed to attack all mobile devices regardless of the platform. 
Moreover, Touchlogger.js is the only works which present the results of the attack for multiple-users modes. By contrast, the results from other works are mainly based on training the classifiers for individual users. In other words, they assume the attacker is able to collect input training data from the victim user before launching the PIN attack. We do not have such an assumption as the training data is obtained from all users in the experiment.
In terms of accuracy, with the exception of \cite{KeyMic}, PINlogger.js generally outperforms other works with an identification rate of 74\% in the first attempt. 
This is a significant success rate (despite that the sampling rate in-browser is much lower than that available in-app) and confirms that the described attack imposes a serious threat to the users' security and privacy.

\section{Why does this vulnerability exist?}
\label{Vul}
Although reports of side channel attacks based on the in-browser access to mobile sensors via JavaScript are relatively recent, similar attacks via in-app access to mobile sensors have been known for years. Yet the problem has not been fixed. 
Here, we discuss the reasons why such a vulnerability has remained unfixed for a long time.

\begin{table*}[t]
\centering
\begin{tabular}{|l|l|l|l|}
\hline
Android & Description &  Unit & W3C def.\\
motion sensors & & &\\
\hline
Accelerometer & Acceleration force  & $m/s^2$& Acceleration \\
& along 3 axes & & with gravity\\
\hline
Gravity & Force of gravity  & $m/s^2$ & NA\\
& along 3 axes&&\\
\hline
Gyroscope & Rate of rotation  & rad/s & Rotation rate\\
&around 3 axes&&\\
\hline
Uncalibrated  & Rate of rotation (no drift & rad/s & NA\\
gyroscope&  compensation), and &&\\
& Estimated drift around 3 axes & rad/s & NA\\
\hline
Linear  & Acceleration force excluding & $m/s^2$ & Acceleration\\
accelerometer & gravity along 3 axes&&\\
\hline
Rotation  & Rotation vector   & Unitless & NA\\
vector& component along 3 axes&&\\
\hline
Step  & Number of user's  & Steps& NA\\
counter & steps since last reboot&&\\
\hline
\end{tabular}
\caption{Motion sensors supported by Android and their corresponding W3C definitions.}
\label{androidmotion}
\end{table*}
\begin{table*}[t]
\centering
\begin{tabular}{|l|l|l|l|}
\hline
Android & Description &  Unit & W3C def.\\
position sensors &&&\\
\hline
Game  & Rotation vector component & Unitless & NA\\
rotation vector & along 3 axes&&\\
\hline
Geomagnetic  & Rotation vector component& Unitless & NA\\
rotation vector&  along 3 axes&&\\
\hline
Geomagnetic  & Geomagnetic field strength & $\mu T$ & NA\\
magnetic field& along 3 axes&&\\
\hline
Uncalibrated & Geomagnetic field strength & $\mu T$ & NA\\
magnetic field & (no hard iron calibration)&&\\
\hline
& and Iron bias estimation & $\mu T$ & NA\\
&along 3 axes&&\\
\hline
Orientation & Angles around 3 axes& Degrees & Orientation\\
\hline
Proximity & Distance from object & cm & NA\\
\hline
\end{tabular}
\caption{Position sensors supported by Android and their corresponding W3C definitions. Note: Orientation sensor was deprecated in Android~2.2 (API Level~8).}
\label{androidposition}
\end{table*}
\subsection{Unmanaged sensors}
In an attempt to explain multiple sensor-related in-app vulnerabilities, Xu et al.\ argue that ``the fundamental problem is that sensing is unmanaged on existing smartphone platforms''~\cite{taplogger}. 
There are multiple in-app side-channel attacks that support this argument, as we discussed in the previous section. 
Our work shows that the problem of in-app access to ``unmanaged sensors'' is now spreading to in-browser access. 
Here we present the ``unmanaged'' motion and orientation sensor case which shows how the technical mismanagement of these sensors causes serious user privacy consequences when it comes to unregulated access to such sensors via JavaScript. 

\textbf{W3C vs. Android.} 
According to W3C specifications~\cite{W3CMotion}, the motion and orientation sensor streams are not raw sensor data, but rather high-level data which are agnostic to the underlying source of information. Common sources of information include gyroscopes, compasses and accelerometers. In Tables~\ref{androidmotion} and \ref{androidposition}, we present 
raw (low-level) and synthesized (high-level) 
motion sensors supported by Android~\cite{AndDev} along with their descriptions and units, as well as their corresponding W3C definitions~\cite{W3CMotion}.

As it can be seen from the tables, different terminologies have been used for describing the same measurements in-app and in-browser. For example, while in-app access uses the raw sensor terminology, i.e., accelerometer, gyroscope, magnetic field, the in-browser access uses synthesized sensor terminology, i.e., motion and orientation \cite{W3CMotion}.
This creates confusion for users (as we will explain later) and developers (as we experienced it ourselves). One of the W3C's specifications on mobile sensors, ``Generic Sensor API'' \cite{Generic}, dedicates a few sections to the issue of naming sensors, and low-level and high-level sensors. It discusses how the terminology for in-browser access has been high-level so far. It also mentions that the low-level use cases are increasingly popular among the developers.      
As stated in this specification: ``The distinction between high-level and low-level sensor types is somewhat arbitrary and the line between the two is often blurred''. And, ``Because the distinction is somewhat blurry, extensions to this specification are encouraged to provide domain-specific definitions of high-level and low-level sensors for the given sensor types they are targeting''.
We believe due to the rapid increase of mobile sensors, it is necessary to come up with a consistent approach.

\subsection{Unknown sensors}
\label{sensorlist}
We believe another contributing factor is that users seem to be less familiar with the relatively newer (and less advertised) sensors such as motion and orientation, as opposed to their immediate familiarity with well-established sensors such as camera and GPS. 
For example, a user has asked this question on a mobile forum: ``... What benefits do having a gyroscope, accelerometer, proximity sensor, digital compass, and barometer offer the user? I understand it has to do with the phone orientation but am unclear in their benefits. Any explanation would be great! Thanks!''\footnote{forums.androidcentral.com/verizon-galaxy-nexus/171482-barometer-accelerometer-how-they-useful.html}.

We design and conduct user studies in this work in order to investigate to what extent are these sensors and their risks known to the users. 

\textbf{List of mobile sensors}.
We prepared a list of different mobile sensors by inspecting the official websites of the latest iOS and Android products, and the specifications that W3C and Android provide for developers. We also added some extra sensors as common sensing mobile hardware which are not covered before.

\begin{itemize}
\item iPhone 6\footnote{apple.com/uk/iphone-6/specs/}: 
Touch ID, Barometer, Three-axis gyro, Accelerometer, Proximity sensor, Ambient light sensor.

\item Nexus 6P\footnote{store.google.com/product/nexus\_6p}:    Fingerprint sensor, Accelerometer, Gyroscope, Barometer, Proximity sensor, Ambient light sensor, Hall sensor, Android Sensor hub.

\item Android \cite{AndDev}: Accelerometer, Ambient temperature, Gravity (software or hardware), Gyroscope, Light, Linear Acceleration (software or hardware), Magnetic Field, Orientation (software), Pressure, proximity, Relative humidity, Rotation vector (Software or Hardware), Temperature. 

\item W3C\footnote{w3.org/2009/dap/} \cite{W3CMotion}: Device orientation (software), Device motion (software), Ambient light, Proximity, Ambient temperature, Humidity, Atmospheric Pressure. 

\item Extra sensors (Common sensing hardware): Wireless technologies (WiFi, Bluetooth, NFC), Camera, Microphone, Touch screen, GPS.    
\end{itemize} 

Unless specified otherwise, all the listed sensors are hardware sensors. We added the last category of the sensors to this list since they indeed sense the device's surrounding although in different ways. However, they are neither counted as sensors in mobile product descriptions, nor in technical specifications. These sensors are often categorised as OS resources~\cite{resources}, and hence different security policies apply to them.

\subsection{User study}
\label{refff}
In this section, we aimed to observe the amount of knowledge that mobile users have about mobile sensors. We prepared a list of sensors based on what we explained above and asked volunteer participants to rate the level of their familiarity with each sensor.
All of our experiments and user studies were approved by Newcastle University's ethical committee.

\subsubsection{Participants}
We recruited 60 participants to take part in this study via different means including mailing lists, social networking, vocational networks, and distributing flyers in different places such as different schools in the university, colleges, local shops, churches and mosques. A sample of our call for participation is available in Appendix~\ref{demoo}.

Among our participants, 28 self-identified themselves as male and 32 as female, from 18 to 67 years old, with a median age of 33.85. 
None of the participants were studying or working in the field of mobile sensor security. Our university participants were from multiple degree programs and levels, and the remaining participants worked in a different range of fields. 
Moreover, our participants owned a wide range of mobile devices, and had been using a smartphone/tablet for 5.6 years on average. 
Our participants were from different countries, and all could speak English. 
We interviewed our participants at a university office and gave each an Amazon voucher (worth \pounds 10) at the end for their participation. 
Details of the interview template can be found in Appendix~\ref{inter}.  

\subsubsection{Study approach}
For a list of 25 different sensors, we used a five-point scale self-rated familiarity questionnaire as used in~\cite{self}: ``I've never heard of this'', ``I've heard of this, but I don't know what this is'', ``I know what this is, but I don't know how this works'', ``I know generally how this works'', and ``I know very well how this works''. 
The list of sensors was randomly ordered for each user to minimize bias. 
In addition, we needed to observe the experiments to make sure users were answering the questions based on their own knowledge in order to avoid the effect of processed answers. Full descriptions of all studies are provided in Appendix \ref{inter}. 

\begin{figure*}[!t]
	\centering
	\includegraphics[scale = 0.25]{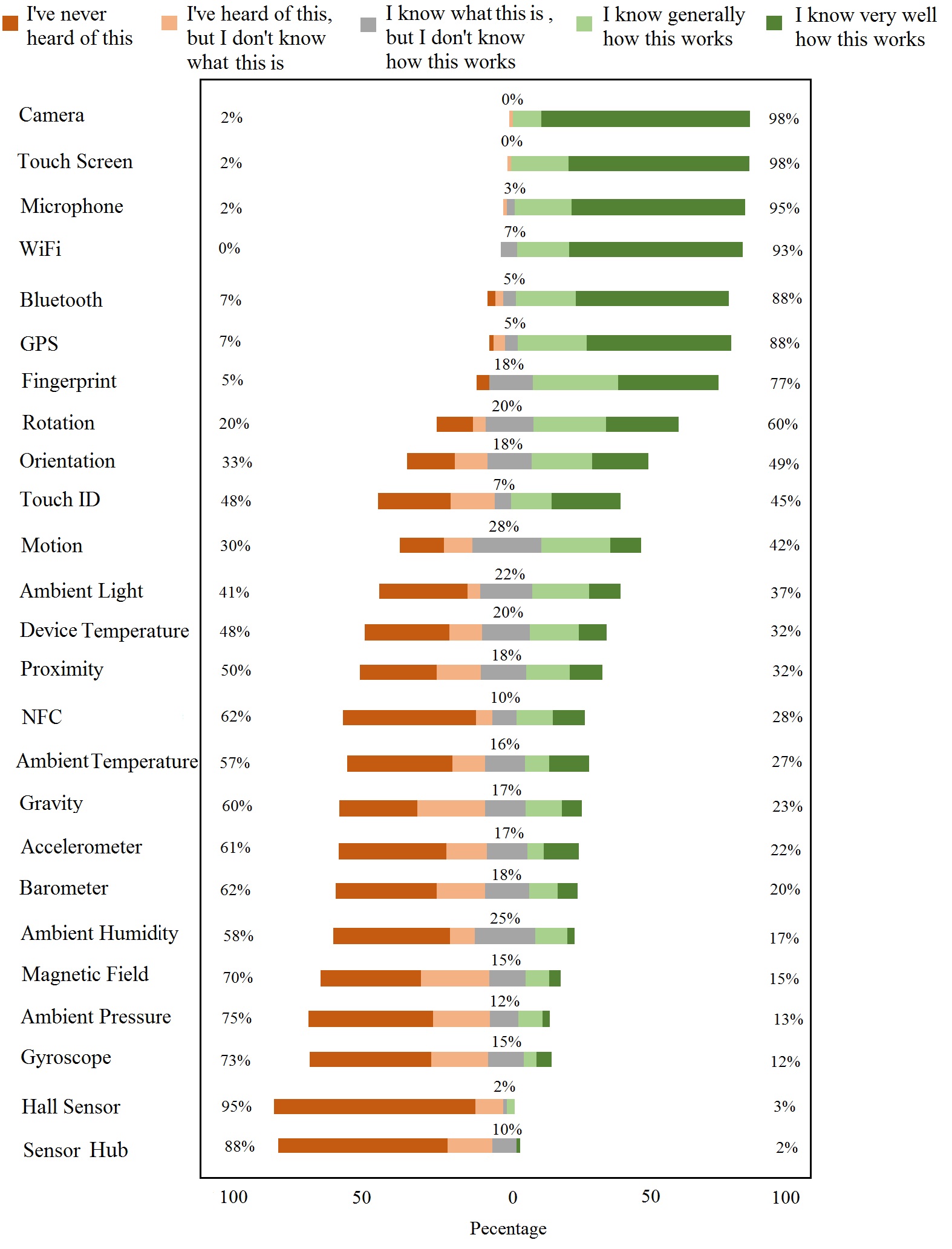}
	\caption{ Level of self-declared knowledge about different mobile sensors.}
	\label{userstudy}
\end{figure*}
\subsubsection{Findings}
Fig.~\ref{userstudy} summarizes the results of this study. 
This figure shows the level of self-declared knowledge about different mobile sensors. The question was: ``To what extent do you know each sensor on a mobile device?'' Sensors are ordered based on the aggregate percentage of participants declaring they know generally or very well how each sensor works. This aggregate percentage is shown on the right hand side. In the case of equal aggregate percentage, the sensor with a bigger share on being known very well by the participants is shown earlier.   
Our participants were generally surprised to hear about some sensors and impressed by the variety. 
As one may expect, newer sensors tend to be less known to the users in comparison to older ones. 
In particular, our participants were generally not familiar with ambient sensors. Although some of our participants knew  the ambient sensors in other contexts (e.g., thermostats used at home), they could not recognise them in the context of a mobile device. 

Low-level hardware sensors such as accelerometer and gyroscope seem to be less known to the users in comparison with high-level software ones such as motion, orientation, and rotation. We suspect that this is partly due to the fact that the high-level sensors are named after their functionalities and can be more immediately related to user activities. 

We also noticed that a few of the participants knew some of the low-level sensors by name but they could not link them to their functionality. For example, one of our participants who knew almost all of the listed sensors (except hall sensor and sensor hub) stated that: ``When I want to buy a mobile [phone], I do a lot of search, that is why I have heard of all of these sensors. But, I know that I do not use them (like accelerometer and gyroscope)''. 

On the other hand, as the functionalities of mobile devices grow, vendors quite naturally turn to promote the software capabilities of their products, instead of introducing the hardware. For example, many mobile devices are recognised for their gesture recognition features by the users, however the same users might not know how these devices provide such a feature. For instance, one of the participants commented on a feature on her smartphone called ``Smart Stay''\footnote{samsung.com/us/support/answer/ANS00035658/234302/SCH-R950TSAUSC} as follows: ``I have another sensor on my phone: Smart Stay. I know how it works, but I don't know which sensors it uses''.

\section{User studies on risk perception of mobile sensors}
In this section, we study the participants' risk perception of mobile sensors. 
There have been several studies on risk perception addressing different aspects of mobile technology.  
Some works discuss the risks that users perceive on smartphone authentication methods such as PINs and patterns \cite{smart}, TouchID and Android face unlock \cite{au}, and implicit authentication \cite{impil}. Other works focus on the privacy risks of certain sensors such as GPS \cite{brothers}. 
In \cite{Raij}, Raji et al.~show users' concerns (on disclosure of selected behaviours and
contexts) about a specific sensor-enabled device called AutoSense\footnote{sites.google.com/site/autosenseproject/}. 
To the best of our knowledge, the research presented in this paper is the first that studies the user risk perception for a comprehensive list of mobile sensors (25 in total). We limit our study to the level of perceived risks users associate with their PINs being discovered by each sensor. The reasons we chose PINs are that first, finding one's PIN is a clear and intuitive security risk, and second, we can put the perceived risk levels in context with respect to the actual risk levels for a number of sensors as described in Table~\ref{compareall}.  

\subsection{Methodology}
For this study, we divide our 60 participants into two groups, and studied the two group separately using two different approaches: within-subject and between-subject.
In the within-subject study, we interviewed 30 participants for all parts of the study. In contrast, in the between-subject study, we interviewed a new group of 30 participants, and we later compared the results with the previous group. By these two approaches, we aim to measure differences (after informing users on descriptions of sensors) within a participant and between participants, respectively.

\subsubsection{Within-subject study}
In this approach, we asked 30 participants to rate the level of risk they perceive for each sensor in regards to revealing their PINs in two phases. In phase one, we gave the same sensor list (randomized for each user). 
We described a specific scenario in which a game app which has access to all these sensors is open in the background and the user is working on his online banking app, entering a PIN. 
We used a self-rated questionnaire with five-point scale answers following the same terminology as used in \cite{Raij}: ``Not concerned'', ``A little concerned'', ``Moderately concerned'', ``Concerned'', and ``Extremely concerned''.
During this phase, we asked the users to rely on the information that they already had about each sensor (see Appendix~\ref{inter} for details).

In the second phase, first we provided the participants with a short description of each sensor and let them know that they can ask further questions until they feel confident that they understand the functionality of all sensors. Participants could use a dictionary on their device to look at the words that were less familiar to them. 
Afterwards, we asked the participants to fill in another copy of the same questionnaire on risk perceptions (details in Appendix~\ref{inter}). Participants could keep the sensor description paper during this phase to refer to it in the case they forgot the description of certain sensors.  

\subsubsection{Between-subject study}
In this study, first we gave the description of the sensors to our second group of 30 participants and similar to previous study we gave them enough time to familiarize themselves with the sensors and to ask as many questions as they wanted until they felt confident about each sensor.  Then, we presented the participants with the questionnaire on risk perceptions (details in Appendix~\ref{inter}). Similar to our previous study, participants could keep the sensor description paper while filling in this questionnaire.  

\begin{figure*}[!t]
	\centering
    \includegraphics[scale = 0.26]{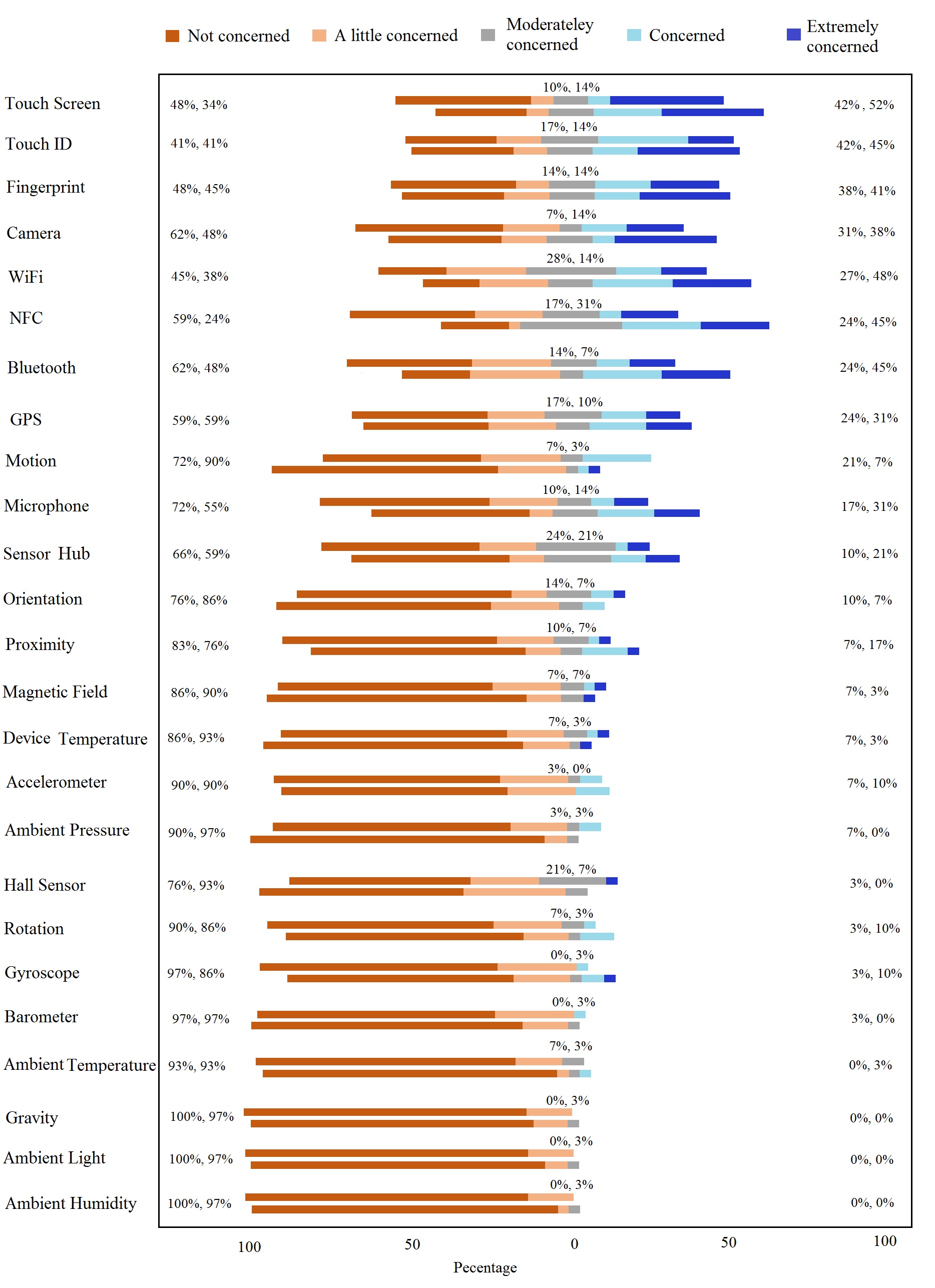}
	\caption{Users' perceived risk for different mobile sensors for within-subject approach.}
	\label{riskper}
\end{figure*}
\begin{figure*}[!t]
	\centering
    \includegraphics[scale = 0.26]{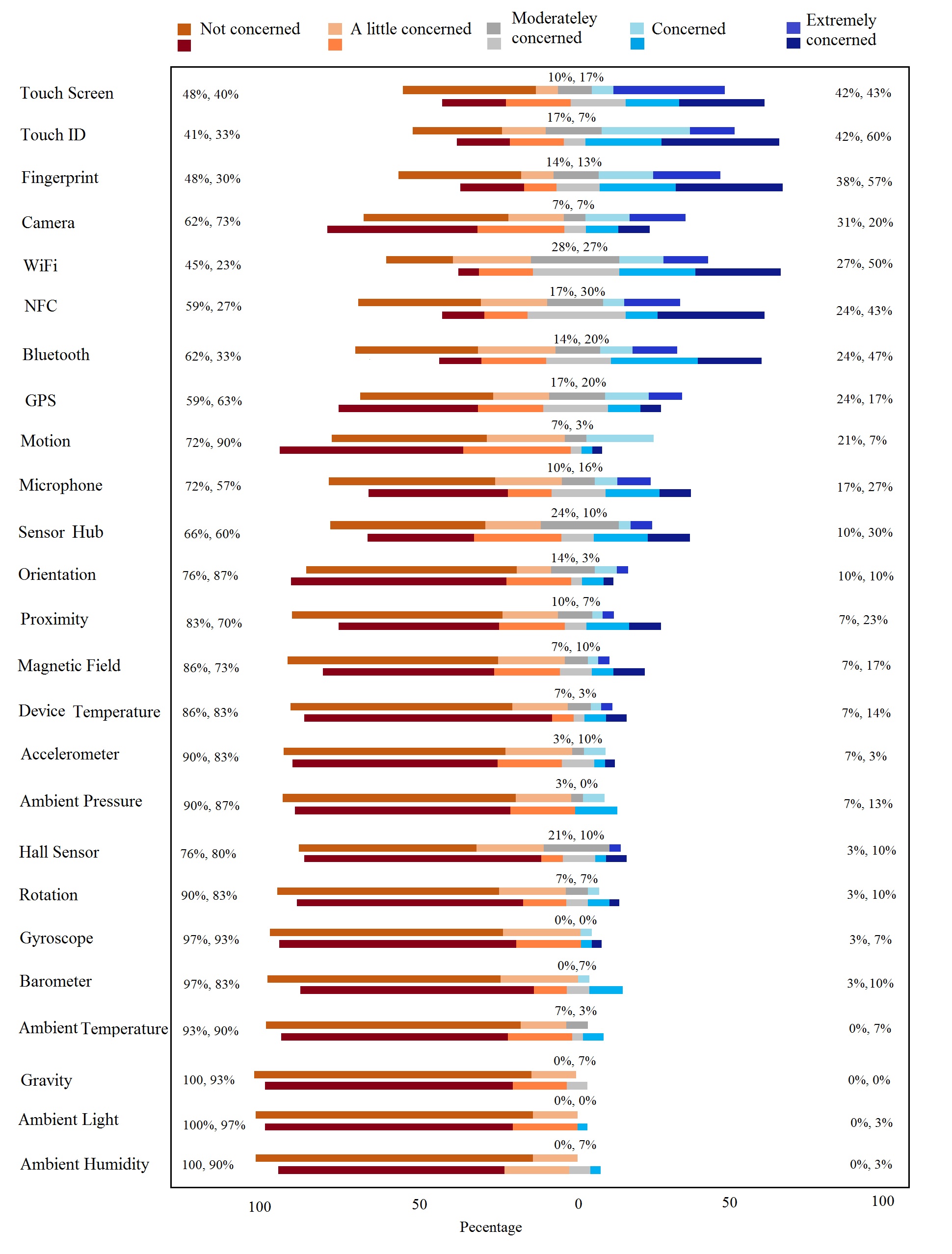}
	\caption{Users' perceived risk for different mobile sensors for between-subject approach.} 
	\label{riskper2}
\end{figure*}
\subsubsection{Intuitive risk perception}
The results of our within-subject study are presented in Fig.~\ref{riskper}. 
These results present the users' perceived risk for different mobile sensors for the same group of users before (top bars) and after (bottom bars) being presented with descriptions of sensors. 
The results of our between-subject study are presented in Fig.~\ref{riskper2}. 
Note that this figure represents the risk perception of group one of our participants before knowing the sensors descriptions, and group two of participants after knowing the sensors descriptions.
For both figures, the question was: ``To what extent are you concerned about each sensor's risk to your PIN?'', sensors are ordered based on the aggregate percentage of participants declaring they are either concerned or extremely concerned about each sensor before seeing the descriptions. This aggregate percentage is the first value presented on the right hand side. In the case of equal aggregate percentage, the sensor with a bigger share on being perceived extremely concerned by the participants is shown earlier.

We make the following observations from the results of the experiment.

\textbf{Touch Screen.} Although our participants rated touch screen as one of the most risky sensors in relation to a PIN discovery scenario, still about half of our participants were either moderately concerned, a little concerned, or not concerned at all. 
Through our conversations with the users, we received some interesting comments, e.g., ``Why any of these sensors should be dangerous on an app while I have officially installed it from a legal place such as Google Play?'', and ``As long as the app with these sensors is in the background, I have no concern at all''. It seems that a more general risk model in relation to mobile devices is affecting the users' perception in regard to the presented PIN discovery threat. This fact can be a topic of research on its own, and is out of the scope of this paper.     

\textbf{Communicational Sensors.} 
One category of the sensors which users are relatively more concerned about includes WiFi, Bluetooth and NFC. For example one of the participants commented that: ``I am not concerned with physical [motion, orientation, accelerometer, etc.]/ environmental [light, pressure, etc.] sensors, but network ones. Hackers might be able to transfer my information and PIN''. These sensors appearing more risky to the users is understandable since we asked them to what extent they were concerned about \emph{each sensor} in regard to the PIN discovery. 

\textbf{Identity-related Sensors.} Another category which has been rated more risky than others contains those sensors which can capture something related to the user's identity i.e. fingerprint, TouchID, GPS, camera, and microphone. Despite that we described a PIN-related scenario, our participants were still concerned about these sensors.   
This was also pointed out by a few participants through the comments. 
For example a user stated: ``..., however, GPS might reveal the location along with the user input PIN that has a risk to reveal who (and where) that PIN belongs to. Also the fingerprint/TouchID might recognize and record the biometrics with the user's PIN''. Some of these sensors such as GPS, fingerprint, and TouchID, however, can not cause the disclosure of PINs on their own. Hence, the concern does not entirely match the actual risk. Similar to the discussion on touch screen, we believe that a more general risk model on mobile technology influences the users to perceive risk on specific threats such as the one we presented to them. 

\textbf{Environmental Sensors.} The level of concern on ambient sensors (humidity, light, pressure, and temperature) is generally low and stays low after the users are provided with the description of the sensors (see Fig.~\ref{riskper}). 
In many cases, our users expressed that they were concerned about these sensors simply because they did not know them: ``[now that I know these sensors,] I am quite certain that movement/environmental sensors would not affect the security of personal id/passwords etc.''.
In fact, researchers have reported that it is possible to infer the user's PIN using the ambient light sensor data~\cite{SkimLight}, although, to our knowledge, exploits of other environmental sensors have not been reported in the literature. 

\textbf{Movement Sensors.} On the sensors related to the movement and the position of the phone (accelerometer, gyroscope, motion, orientation, and rotation), the users display varying levels of the risk perceptions. In some cases they are slightly more concerned, but in others they are less concerned once they know the functionality. Some of our users stated that since they did not know these sensors, they were not concerned at all, but others were more concerned when they were faced with new sensors. Overall, knowing, or not knowing these sensors has not affected the perceived risk level significantly, and they were rated generally low in both cases. 

\textbf{Motion and Orientation Sensors.} 
The sensors which we used in our attack, namely orientation, rotation, and motion, have not been generally scored high for their risk in revealing PINs. 
Users do not seem to be able to relate the risk of these sensors to the disclosure of their PINs, despite that they seem to have an average general understanding about how they work. 
On hardware sensors such as accelerometer and gyroscope, the risk perception seems to be even lower. 
A few comments include: ``In my everyday life, I don't even think about these [movement] sensors and their security. There is nothing on the news about their risk'', and ``I have never been thinking about these [movement] sensors and I have not heard about their risk''. On the other hand, some of the participants expressed more concerns for sensors that they were familiar with, as one wrote, ``You always hear about privacy stuff for example on Facebook when you put your location or pictures''. Similarly, it seems that having a previous risk model is a factor that might explain the correlation between the user's knowledge and their perceived risk. 

\section{Discussions} 
\subsection{General knowledge versus risk perception}
Figs.~\ref{userstudy} and \ref{riskper} suggest that there may be a correlation between the relative level of knowledge users have about sensors and the relative level of risk they perceive from them. 
We confirm our observation of correlation using Spearman's rank-order correlation measure. As it can be seen in Table \ref{spearman}, we present the Spearman's correlation between the comparative knowledge and the perceived risk about different sensors for different participants' dataset: group one before being presented with the sensor descriptions, group one after sensor description, group two after sensor descriptions, and finally groups one and two after being presented with the sensor descriptions. 

For each participants' dataset, the sensors are separately ranked based on the level that the users are familiar with them, similar to Figure \ref{userstudy}. Accordingly, the levels of concern are ranked too. The Spearman's correlation equation has been applied on these ranks for each group separately. 

For example, the Spearman's correlation between the comparative knowledge (median: ``I know what this is, but I don't know how this works'', IQR\footnote{interquartile range}: ``I've never heard of this'' -- ``I know very well how this works'') 
and the perceived risk about different sensors for group one (median: ``Not concerned'', IQR: ``Not concerned'' -- ``A little concerned'') before knowing the sensor descriptions is $r = 0.61$ ($p<0.05$). 

As it can be seen, these results support that the more the users know about these sensors, the more concern they express about the risk of the sensors revealing PINs. 
We acknowledge that other methods of ranking the results, e.g. using median, produce slightly different final rankings. 
However, given the high confidence level of the above test, we expect the correlation to be supported if other methods of ranking are used. 

\begin{table}[t]
\centering
\begin{tabular}{|l|l|l|}
\hline
Participants' & Status & Spearman's \\
 dataset &  & correlation\\
\hline
Group 1& Before sensor desc.& 0.61\\
Group 1& After sensor desc.& 0.61 \\
Group 2& After sensor desc.& 0.48\\
Groups 1 and 2 & After sensor desc. & 0.58 \\
\hline
\end{tabular}
\caption{Spearman's correlation between the comparative knowledge and the perceived risk about different sensors.}
\label{spearman}
\end{table}

Assuming that customer demand drives better security designs, the above correlation may explain why sensors that are newer to the market have not been considered as OS resources and consequently have not been subject to similar strict access control policies. 

\subsection{Perceived risk vs the actual risk}
We are specifically interested in the users' relative risk perception of sensors in revealing their PINs in comparison to the actual relative risk level of these sensors. We list the results reported in the literature in Table~\ref{compareall} for the following sensors: light, camera, microphone, gyroscope, motion, and orientation. 
Fig.~\ref{riskper} shows that users generally have expressed more concern about sensors such as camera and microphone than accelerometer, gyroscope, orientation, and motion. This does not match the actual risk levels since the latter sensors allow PIN recovery with higher accuracy as we have shown in Section \ref{Eva}. 
When asked after filling the questionnaire, most participants could not come up with realistic attack scenarios using camera and microphone. For microphone, some users thought they might say the PIN out loud. For camera, a few of our participants thought face recognition might be used to recover the PIN, hence they rated camera's risk to their PINs high. One user thought the camera might capture the reflection of the entered PIN in her glasses. 

Among our participants, one mentioned but described doubt about motion, orientation, accelerometer, and gyroscope being able to record the shakes of the mobile phone while entering a PIN after they saw the sensor descriptions: ``I feel those positional sensors might be able to reveal something about my activities, for example if I open my banking app or enter my PIN. But it is extremely hard for different users, and when working with different hands and positions''.  
This participant expressed only ``a little concern'' about them, stating that: ``..., and by little concern, I mean extremely little concern". One of our participants was completely familiar with these attacks and in fact had read some related papers. This user was ``extremely concerned''. Other users who rated these sensors risky in general, said they were generally concerned about different sensors. One commented: ``I can not think of any particular situation in which these sensors can steal my PIN, but the hackers can do everything these days.'' 

\subsection{Possible Solutions}
\label{Sol}
In this section, we discuss the current academic and industrial countermeasures to mitigate sensor-based attacks. 

\subsubsection{Academic approach} 
Different solutions to address the in-app access attacks have been suggested in the literature: e.g.,  
restricting the sensor to one app, reducing the sampling rate, temporal pause of the sensor on sensitive entries such as keyboard, rearranging keyboard for password entrance, asking for explicit permission from the user, ranking apps based on their similarities to malware, and obfuscating anomalies in sensor data \cite{KeyMic,Tapprints2,SkimLight,taplogger,PINCamera,Speech:Gyr,Tapprints,accessory,Finger,Finger2}. 
However, after many years of research on showing the serious security risks of sensors such as accelerometer and gyroscope, none of the major mobile platforms have revised their in-app access policy.

We believe that the risks of unmanaged sensors on mobile phones, specially through JavaScript code, are not known very well yet. More specifically, many OS/app level solutions such as asking for permissions at the installation time, or malware detection approaches would not work in the context of a web attack. 
In our previous work \cite{Mehrnezhad}, we suggested to apply the same security policies as those for camera, microphone, and GPS for the motion and orientation sensors. Our suggestion was to set a multi-layer access control system on the OS and browser levels. However, the usability and effectiveness of this solution are arguable. 
First, asking too many permissions from the user for different sensors might not be usable. 
Furthermore, for some basic use cases such as gesture recognition to clear a web form, or adjusting the screen from portrait to landscape, it might not make sense to ask for user permission for every website. 
Second, with the increase of the number of sensors accessible through mobile browsers, this approach might not be effective due to the classic problem of sidestepping the security procedure by users when it is too much of a burden~\cite{attention}. 
As stated by one of our participants: ``I don't mind these sensors being risky anyway. I don't even review the permission list. I have no other choice to be able to use the app''.
Moreover, as we have shown in Section~\ref{Vul}, users generally do not understand the implications of these sensors on discovering their PINs for example, even though they know how these sensors work. Hence, such an approach might not be effective in practice.

\subsubsection{Industrial approach} 

\textbf{W3C Device Orientation Event Specification.}
There is no Security and Privacy section in the latest official W3C Working Draft Document on Device Orientation Event~\cite{W3CMotion}. However, at the time of writing this paper, a new version of the W3C specification is being drafted, which includes a new section on security and privacy issues related to mobile sensors\footnote{w3c.github.io/deviceorientation/spec-source-orientation.html}, as suggested by us in~\cite{Mehrnezhad}.  
The authors working on the revision of the W3C specification point out the problem of fingerprinting mobile devices~\cite{fin}, and touch action recovery~\cite{Mehrnezhad} through these sensors, and suggest the following mitigations: 
\begin{itemize}
\item ``Do not fire events when the page where they were registered on is not visible or has been backgrounded.''
\item ``Fire events only on the top-level browsing context or same-origin nested iframes.''
\item ``Limit the frequency of events (typically 60~Hz seems to be sufficient).''
\end{itemize}

We believe that these measures may be too restrictive in blocking useful functionalities. For example, imagine a user consciously running a web program in the browser to monitor his daily physical activities such as walking and running. This program needs to continue to have access to the motion and orientation sensor data when the user is working on another tab or minimizes the browser. One might argue that such a program should be available as an app instead, hence the use case is not valid. However, it is expected that the boundary between installed apps and embedded JavaScript programs in the browser will gradually diminish~\cite{webvnative}. 

\textbf{Mobile browsers.} 
As we showed in \cite{Mehrnezhad}, browsers and mobile operating systems behave differently on providing access to sensors. Some allow access only on the active webpage and any embedded iframes (although with different origins), some allow access to other tabs, when browser is minimized, or even when the phone is locked. Hence, there is not a consistent approach across all browsers and mobile platforms. 
Reducing the frequency rate has been applied to all well-known browsers at the moment~\cite{Mehrnezhad}. For instance, Chrome reduced the sensor readings from 200~Hz to 60~Hz due to security concerns\footnote{bugs.chromium.org/p/chromium/issues/detail?id=421691}. 
However, our attack shows that security risks are still present even at lower frequencies. 
iOS and Android limit the maximum frequency rate of some sensors such as Gyroscope to 100~Hz and 200~Hz, respectively. It is expected that these frequencies will increase on mobile OSs in the near future and in-browser access is no exception. 
In fact, current mobile gyroscopes support much higher sampling frequencies, e.g., up to 800~Hz by STMicroelectronics (on Apple products), and up to 8000~Hz by InvenSense (on the Google Nexus range)~\cite{Speech:Gyr}. 
With higher frequencies available, attacks such as ours can perform better in the future if adequate security countermeasures are not applied. 

Following our report of the issue to Mozilla, starting from version 46 (released in April 2016), Firefox restricts JavaScript access to motion and orientation sensors to only top-level documents and same-origin iframes\footnote{mozilla.org/en-US/security/advisories/mfsa2016-43/}. 
In the latest Apple Security Updates for iOS~9.3 (released in March 2016), Safari took a similar countermeasure by ``suspending the availability of this [motion and orientation] data when the web view is hidden''\footnote{support.apple.com/en-gb/HT206166}. 
However, we believe the implemented countermeasures should only serve as a temporary fix rather than the ultimate solution. 
In particular, we are concerned that it has the drawback of prohibiting potentially useful web applications in the future.
For example, a web page running a fitness program has a legitimate reason to access the motion sensors even when the web page view is hidden. 
However, this is no longer possible in the new versions of Firefox and Safari. 
Our concern is confirmed by members in the Google Chromium team\footnote{bugs.chromium.org/p/chromium/issues/detail?id=523320}, who also believe that the issue remains unresolved.

\subsection{Biometric sensors}
As we explained in section \ref{sensorlist}, there exist around 25 different sensors on mobile platforms. They include communicational  sensors such as WiFi, environmental sensors such as ambient light, movement sensors such as motion and orientation, and biometric sensors such as Fingerprint. Here we specifically discuss biometric sensors since they are highly related to the individuals' identity. 

After decades of working on password, it seems that people still cannot remember strong passwords. 
Biometrics have been offered to users as an effective authentication mechanism. Examples include TouchID and Fingerprint sensors on iOS and Android devices respectively. But the biometric-based authentication is not limited to mobile devices only. For example, when paying with iPhone contactlessly, you need to rest your finger on TouchID and hold your iPhone in close proximity to the contactless reader until the task is finished. Furthermore, since many banks have already moved their services to mobile platforms, they benefit from the biometrics sensors available on mobile devices, say for implementing 2-factor authentication. As an example, in addition to user name and passwords, HSBC authenticates their customers through TouchID\footnote{us.hsbc.com/1/2/home/personal-banking/pib/mobile/touchid} and voice ID\footnote{hsbc.co.uk/1/2/contact-and-support/banking-made-easy/voice-id}. Another example is \textit{Smile to Pay} facial recognition app\footnote{brandchannel.com/2015/03/16/alibaba-demos-smile-to-pay-facial-recognition-app/} where deep learning is applied to overcome the difficulty of face authentication when the face photo is not in the normal form. Recently Yahoo has also introduced its ear-based smartphone identification system\footnote{bbc.co.uk/news/technology-32498222}.

On the other hand, our findings show that mobile users are relatively concerned with identity-related or biometric sensors. However, we discussed that these sensors are not necessarily the most risky ones to PINs in practice. As we mentioned earlier, we believe that this might be the influence of a more general risk model that the users have on mobile technology. We believe that this is an important research topic and requires further studies. 

\subsection{Limitations}
We consider this work a pilot study that explores user risk perception on a comprehensive list of mobile sensors. We envisage the following future work to address these limitations and expand this work: 
\begin{itemize}
\item \textit{More Participants}: We performed our user studies on a set of users who were recruited from a wide range of backgrounds. Yet the number of the participants is limited. A larger set of participants will improve the confidence in the results. With a large and diverse set of participants, we can also study the effect of demographic factors on perceived risk. 

\item \textit{Other Risks}: We studied the perceived risk on PINs as a serious and immediate risk to users' security. The study can be expanded by studying users' risk perception on other issues such as attackers discovering phone call timing, physical activities, or shopping habits. 

\item \textit{Other Types of Access}: When interviewing our participants, we presented them with a scenario involving a game app which is installed on their smartphone. This only covers the in-app access to sensors. However, people might express different risk levels for other types of access, e.g., in-browser access. This needs further investigation. 

\item \textit{Issues with Training Users}. We decided to provide our participants with a short description of each sensor's functionality (details in Appendix~\ref{inter}, part 3). Furthermore, the participants were given the chance to ask as many questions as they wanted to fully understand the functionality of each sensor. This might not be the most effective way to inform users about sensors since some descriptions might seem too technical (and hence not fully understandable) to some users. How to inform users in an effective way is a complex topic of research which can be explored in the future. 

\end{itemize}

\section{Conclusion}
\label{Con}
In this paper, we introduced PINlogger.js, a web-based program which reveals users' PINs by recording the mobile device's orientation and motion sensor data through JavaScript code. 
Access to mobile sensor data via JavaScript is limited to only a few sensors at the moment. This will probably expand in the future, specially with the rapid development of sensor-enabled devices in the Internet of things (IoT). 

We also showed that users do not generally perceive a high risk about such sensors being able to steal their PINs. Furthermore, we showed that people are not even generally knowledgeable about these sensors on mobile devices. Accordingly, we discussed the complexity of designing a usable and secure solution to prevent the proposed attacks.
Hence, designing a general mechanism for secure and usable sensor data management remains a crucial open problem for future research.

Many of the suggested academic solutions either have not been applied by the industry as a practical solution, or have failed. Given the results in our user studies, designing a practical solution for this problem does not seem to be straightforward.  A combination of different approaches might help researchers devise a usable and secure solution. Having control on granting access \emph{before} opening a website and \emph{during} working with it, in combination with a smart notification feature in the browser would probably achieve a balance between security and usability. 
Users should also have control on reviewing, updating and deleting these data, if stored by the website or shared with a third party \emph{afterwards}.  
Solutions such as Taintroid~\cite{rod}, a tracking app for monitoring sources of sensitive data on a mobile which has been applied for GPS in~\cite{brothers} could be helpful. After all, it seems that an extensive study is required toward designing a permission framework which is usable and secure at the same time. Such research is a very important usable security and privacy topic to be explored further in the future. 

\section*{Acknowledgements}
We would like to thank professor Angela Saase for her inspirational speech at the annual Privacy Enhancing Technologies Symposium 2016 (PETS2016), which has influenced some parts of this paper. 
We would like to thank Dr. Kovila Coopamootoo from Newcastle University for her constructive feedback on designing the user studies of this paper. 
We also would like to thank the voluntary participants who contributed to our data collection and user studies.
The last three authors are supported by ERC Starting Grant No.~306994. 

This paper is based on: “Stealing PINs via Mobile Sensors: Actual Risk versus User Perception.”, by Maryam Mehrnezhad, Ehsan Toreini, Siamak F. Shahandashti, Feng Hao which appeared in the Proceedings of EuroUSEC, 2016.
\bibliographystyle{abbrv}

\appendix
\section{Call for Participation Flyer and Participant Demographics} 
\label{demoo}
In this section, we present the participants demographics in details and the flyers that we used for call for participation of our user studies. 
\begin{figure*}[!t]
	\centering
	\includegraphics[scale = 0.35]{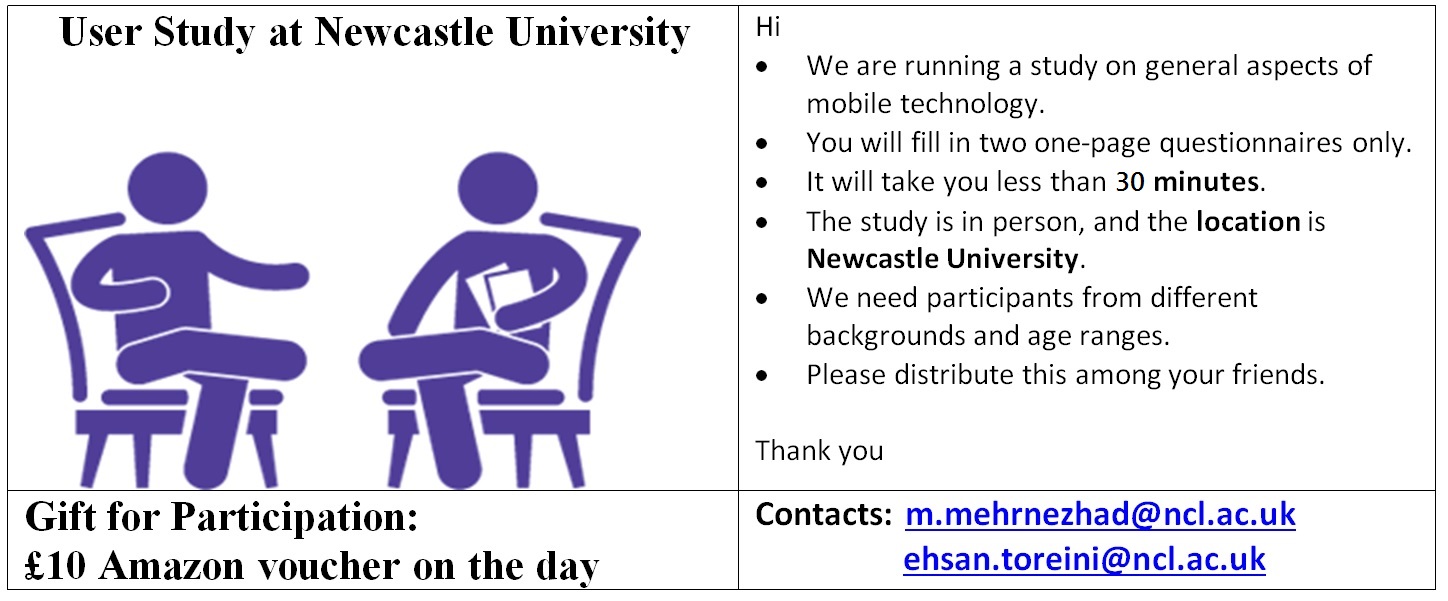}
	\caption{Sample of flyer distributed for participant recruitment.}
	\label{flyer}
\end{figure*}
\begin{table*}[t]
\centering
\begin{tabular}{|l|l|l|l||l|l|l|l|}
\hline
Sex & Age & Job/Background& Mobile (y)& Sex & Age & Job/Background& Mobile (y)\\
\hline
f& 23& Civil Eng.& Nokia (0)&
f& 27& Teacher& HTC(3)\\
f& 28& Customer Support& HTC (2)& 
m& 30& Services& iPhone (4)\\
f& 22& Media & Sony (3)& 
m& 26& Computer & Samsung (7)\\
m& 43& IT& iPhone (9)& 
m& 30& Teacher& Blackberry (7)\\
f& 27& Media & iPhone (9)& 
m& 52& Nanotechnology& Nokia (0)\\
m& 18& Mathematics& Samsung (3)& 
m& 41& Nanotechnology& HTC (10)\\
f& 30& Management& iPhone (7)& 
m& 47& Lecturer& Samsung (2)\\
m& 22& Medical& iPhone (10)& 
f& 39& Physics& iPhone (4)\\
f& 27& Human Mgmt.& Huawei (9)& 
f& 31& Biology & Samsung (10)\\
f& 21& Literature& Samsung (4)& 
m& 39& Student& iPhone (6)\\
m& 35& Media & Samsung (6)& 
f& 30& Civil Eng.& iPhone (5)\\
f& 20& Languages& Samsung (3)& 
m& 20& Student& Samsung (4)\\
f& 59& Services& iPhone (3)& 
f& 52& Admin& Samsung (3)\\
m& 40& IT& LG (7)& 
f& 30& Admin& Samsung (5)\\
m& 21& Biomedical& Samsung (4)&
f& 58& Admin &iPhone (12)\\
f& 22& Biomedical& OnePlus (6)&
f& 44& Admin& Samsung (3)\\
m& 30& Civil Eng.& Samsung (3)&
f& 27& Student& Motorola (5)\\
m& 29& Geodesy& Samsung (7)&
f& 47& Services& iPhone (5)\\
m& 28& Medical& Sony (5)&
m& 67& Teacher& Nokia (0)\\
f& 38& Computer& Samsung (5)&
m& 23& Student& Nexus (5)\\
f& 30& Animation& iPhone (9)&
m& 46& Cable Maker& iPhone (5)\\
f& 56& Business Mgmt.& iPhone (11)&
m& 35& Services& Samsung (5)\\
f& 29& Admin& Samsung (5)&
f& 39& Admin& iPhone(5)\\
f& 30& Admin& Samsung (6)&
f& 24& Student& Gionee (3)\\
m& 47& Driving Instructor& Sony (11)&
f& 34& Education & iPhone  (4)\\
f& 28& Admin& Motorola (7)&
m& 32& Student& OnePlus (6)\\
m& 40& Education& LG (5)&
f& 37& Researcher& Honor (3)\\
m& 32& Computer & iPhone (6)&
m& 33& Industrial Mgmt.& iPhone(12)\\
f& 25& Law& HTC (3)&
f& 33& Mathematics& Samsung (3)\\
m& 30& Student& Nexus (5)&
m& 27& Student& iPhone (18)\\
\hline
\end{tabular}
\caption{Participants' self-reported demographics in the two studies, (y) indicates the years of owning a smartphone}
\label{dem}
\end{table*}
\section{Interview Script}
\label{inter}
Hi. Thanks very much for contributing to our study. In this interview, we will ask you to fill in a few questionnaires about mobile sensors such as GPS, camera, light, motion and orientation.  
You are encouraged to think out loud as you go through, and please feel free to provide any comments during the interview. There is no right or wrong answer, and our purpose is to evaluate the mobile sensors, not you. 
Everything about this interview is anonymous. Please provide some information about yourself in Table \ref{infoo}.
\begin {table}[h]
\begin{center}
 \begin{tabular} { | l | l | }
 \hline
 Age & \\ \hline
 Gender & \\ \hline
 Profession/ background (optional) & \\ \hline
 1st language (optional)& \\ \hline
 Mobile device   & \\ \hline
 Duration of owning a smartphone/tablet   & \\ 
  \hline
\end{tabular}
\caption{Demography}
\label{infoo}
\end{center}
\end {table}
\section*{Part One}
A list of multiple mobile sensors is presented below. 
To what extent do you know each sensor on a mobile device? Please rate them in the table (Table \ref{know} was used).
\begin {table*}[t]
\begin{center}
\begin{tabular}{ |l||l|l|l|l|l|} 
 \hline
 Sensor& 
 I've never & 
 I've heard & 
 I know what & 
 I know  & 
 I know \\ 
 
  & heard of this  
  & of this but I 
  & this is but I
  & generally  
  & very well \\ 
  
  &
  & 
  don't know& 
  don't know&
  how this  &
  how this 
  \\
  
  &
  &
  what this is
  & how this works
  & works
  & works
  \\\hline
Bluetooth&&&&&\\ \hline
Gyroscope&&&&&\\ \hline
GPS&&&&&\\ \hline
Sensor Hub&&&&&\\ \hline
Ambient Temperature&&&&&\\ 
\hline
Accelerometer&&&&&\\ \hline
Magnetic Field&&&&&\\ \hline
Motion&&&&&\\ \hline
Fingerprint&&&&&\\ \hline
Orientation&&&&&\\ \hline
Proximity &&&&&\\ \hline
Ambient Pressure&&&&&\\ 
\hline
Hall Sensor&&&&&\\ \hline
Rotation&&&&&\\ \hline
Touch Screen&&&&&\\ \hline
Camera&&&&&\\  \hline
TouchID &&&&&\\ \hline
Barometer&&&&&\\ \hline
Gravity&&&&&\\ \hline
Microphone&&&&&\\ \hline
Ambient Humidity&&&&&\\
\hline
WiFi&&&&&\\ \hline
Ambient Light&&&&&\\ \hline
NFC&&&&&\\ \hline
Device Temperature&&&&&\\
\hline
\end{tabular}
\caption{This form was used for part one}
\label{know}
\end{center}
\end {table*}
\section*{Part Two}
Imagine that you own a smartphone which is equipped with all these sensors. Consider this scenario: you have opened a game app which can have access to all mobile sensors. 
You leave the game app open in the background, and open your banking app which requires you to enter your PIN. 

Do you think any of these sensors can help the game app discover your entered PIN? 
To what extent are you concerned about each sensor's risk to your PIN? 
Please rate them in the table (Table \ref{Sec1} was used). 
In this section, please only rely on the knowledge you already have about the sensors, and if you do not know some of them, describe your feeling of security about them.  
\begin {table*}[t]
\begin{center}
\begin{tabular}{ |l||l|l|l|l|l|} 
 \hline 
 \multicolumn{6}{c}{Risk to PIN}\\ \hline
 &Not & A little &Moderately &&Extremely\\
  Sensor&Concerned &Concerned&Concerned&Concerned&Concerned\\\hline

Bluetooth&&&&&\\ \hline
Gyroscope&&&&&\\ \hline
GPS&&&&&\\ \hline
Sensor Hub&&&&&\\ \hline
Ambient Temperature&&&&&\\ \hline
Accelerometer&&&&&\\ \hline
Magnetic Field&&&&&\\ \hline
Motion&&&&&\\ \hline
Fingerprint&&&&&\\ \hline
Orientation&&&&&\\ \hline
Proximity &&&&&\\ \hline
Ambient Pressure&&&&&\\ \hline
Hall Sensor&&&&&\\ \hline
Rotation&&&&&\\ \hline
Touch Screen&&&&&\\ \hline
Camera&&&&&\\  \hline
TouchID &&&&&\\ \hline
Barometer&&&&&\\ \hline
Gravity&&&&&\\ \hline
Microphone&&&&&\\ \hline
Ambient Humidity&&&&&\\ \hline
WiFi&&&&&\\ \hline
Ambient Light&&&&&\\ \hline
NFC&&&&&\\ \hline
Device Temperature&&&&&\\\hline          
\end{tabular} 
\caption{This form was used for parts two and three}
\label{Sec1}
\end{center}
\end {table*}
\section*{Part Three}
\label{part3}
Let us explain each sensor here:
\begin{itemize}
\item GPS: identifies the real-world geographic location. 
\item Camera, Microphone: capture pictures/videos and voice, respectively.
\item Fingerprint, TouchID: scans the fingerprint. 
\item Touch Screen: enables the user to interact directly with the display by physically touching it. 
\item WiFi: is a wireless technology that allows the device to connect to a network.
\item Bluetooth: is a wireless technology for exchanging data over short distances.
\item NFC (Near Filed Communication): is a wireless technology for exchanging data over shorter distances (less than 10 cm) for purposes such as contacless payment.  
\item Proximity: measures the distance of objects from the touch screen. 
\item Ambient Light: measures the light level in the environment of the device.
\item Ambient Pressure (Barometer), Ambient Humidity, and Ambient Temperature: measure the air pressure, humidity, and temperature in the environment of the device, respectively.
\item Device Temperature: measures the temperature of the device. 
\item Gravity: measures the force of gravity.
\item Magnetic Field: reports the ambient magnetic field intensity around the device.
\item Hall sensor: produces voltage based on the magnetic field.
\item Accelerometer: measures the acceleration of the device movement or vibration.
\item Rotation: reports how much and in what direction the device is rotated. 
\item Gyroscope: estimates the rotation rate of the device. 
\item Motion: measures the acceleration and the rotation of the device.  
\item Orientation: reports the physical angle that the device is held in.
\item Sensor Hub: is an activity recognition sensor and its purpose is to monitor the device's movement.
\end{itemize}

\noindent 
Please feel free to ask us about any of these sensors for more information. 

Now that you have more knowledge about the sensors, let us describe the same scenario here again.  Imagine that you own a smartphone which is equipped with all these sensors. You have opened a game app which can have access to all mobile sensors. 
You leave the game app open in the background, and open your banking app which requires you to enter your PIN. 

Do you think any of these sensors can help the game app to discover your entered PIN? To what extent are you concerned about each sensor's risk to your PIN? Please rate them in the table (Table \ref{Sec1} was used).
In this part, please make sure that you know the functionality of all the sensors. If you are unsure, please have another look at the descriptions, or ask us about them. 

Thanks very much for taking part in this study. Please leave any extra comment here.

An Amazon voucher and a business card are in this envelope. Please contact us if you have any questions about this interview, or are interested in the results of this study. 

\end{document}